\documentclass{emulateapj}

\begin{document}

\shortauthors{Berdyugina et al.}
\shorttitle{First Detection of a Strong Magnetic Field in a Bursty Brown Dwarf}

\title{First Detection of a Strong Magnetic Field on a Bursty Brown Dwarf:\\
Puzzle Solved}

\author{S. V. Berdyugina\altaffilmark{1,2}, 
D. M. Harrington\altaffilmark{1,2}, 
O. Kuzmychov\altaffilmark{1}, 
J. R. Kuhn\altaffilmark{2},
G. Hallinan\altaffilmark{3}, 
A. F. Kowalski\altaffilmark{4,5},
S. L. Hawley\altaffilmark{6}
}

\altaffiltext{1}{Kiepenheuer Institut f\"ur Sonnenphysik, Sch\"oneckstrasse 6, D-79104 Freiburg, Germany; sveta@kis.uni-freiburg.de }

\altaffiltext{2}{Institute for Astronomy, University of Hawaii, 2680 Woodlawn Drive, Honolulu, 96822-1897 HA, USA}

\altaffiltext{3}{California Institute of Technology, 1200 East California Boulevard, Pasadena, California 91125, USA}

\altaffiltext{4}{Department of Astrophysical and Planetary Sciences, University of Colorado Boulder, 2000 Colorado Ave, Boulder, CO 80305, USA.}

\altaffiltext{5}{National Solar Observatory, University of Colorado Boulder, 3665 Discovery Drive, Boulder, CO 80303, USA.}

\altaffiltext{6}{Department of Astronomy, University of Washington, 15th Avenue NE, Seattle, WA 98195, USA}

\begin{abstract}
We report the first direct detection of a strong, 5\,kG magnetic field on the surface 
of an active brown dwarf. LSR J1835+3259 is an M8.5 dwarf exhibiting transient radio 
and optical emission bursts modulated by fast rotation. We have detected the surface 
magnetic field as circularly polarized signatures in the 819\,nm sodium lines when 
an active emission region faced the Earth. Modeling Stokes profiles of these lines 
reveals the effective temperature of 2800\,K and log gravity acceleration of 4.5. 
These parameters place LSR J1835+3259 on evolutionary tracks as a young brown dwarf 
with the mass of 55$\pm$4\,M$_{\rm J}$ and age of 22$\pm$4\,Myr. Its magnetic field 
is at least 5.1\,kG and covers at least 11\%\ of the visible hemisphere. 
The active region topology recovered using line profile inversions comprises 
hot plasma loops with a vertical stratification of optical and radio emission sources. 
These loops rotate with the dwarf in and out of view causing periodic emission bursts. 
The magnetic field is detected at the base of the loops. This is the first time 
that we can quantitatively associate brown dwarf non-thermal bursts with a strong, 
5 kG surface magnetic field and solve the puzzle of their driving mechanism. 
This is also the coolest known dwarf with such a strong surface magnetic field. 
The young age of LSR J1835+3259 implies that it may still maintain a disk, 
which may facilitate bursts via magnetospheric accretion, like in higher-mass 
T Tau-type stars. Our results pave a path toward magnetic studies of brown dwarfs 
and hot Jupiters.
\end{abstract}


\section{Introduction}

Brown dwarfs are intermediate between stars and planets. Because of their low masses,
they are incapable of hydrogen fusion in the core, so they radiate energy due to deuterium burning, 
slow gravitational collapse, and thermal cooling. 
There is growing evidence that brown dwarfs may possess rather strong magnetic fields, 
similar to active, early M-type red dwarf stars of larger masses and higher temperatures 
\citep{jkv00}. It was found that non-radiative heating 
of plasma, evidenced by chromospheric hydrogen emission and coronal X-ray radiation, 
sharply decline in brown dwarfs \citep{neu99,moh03,pre05,gro07}.
Whether this reduction implies a drop in the surface 
magnetic field strength remains a puzzle.

One clue comes from extremely energetic flares that 
are detected in UV and X-ray radiation as well as in the optical hydrogen Balmer emission, 
providing an indication for magnetic reconnection events \citep{sch07}.  Other evidence 
for magnetic activity in ultra-cool dwarfs is the presence of both quiescent and flaring 
non-thermal radio emission \citep[e.g.,][]{berg01,berg06,hal06,hal07,ant07}. 
A number of them exhibit transient events with periodic radio bursts and aurora-like, 
hydrogen and metal line optical emission \citep{hal07,hal15} but no X-ray corona as 
in warmer, more massive, active red dwarf stars \citep{neu99,pre05,gro07}. 
The radio luminosity of some ultra-cool M dwarfs and brown dwarfs 
is even higher than that of warmer, earlier-type active M dwarfs, and it continues 
to increase beyond spectral type M8 \citep{berg06}, 
the border where both cool, old low-mass stars and warm, young brown dwarfs can be found. 
This peculiar behavior suggests that the “classical” hot stellar chromosphere and 
corona change dramatically across this mass range.

Several very-low-mass ultra-cool stars and brown dwarfs (classes M8 to T6) showing quiet non-thermal 
radio emission were found to be sources of transient but periodic and highly circularly polarized 
bursts at a few GHz, with periods of 2--3 hr, which were associated with rotational periods of dwarfs 
\citep{berg05,hal06,hal07}. 
It was suggested \citep{hal06} that these bursts may be produced at polar regions of a large-scale 
magnetic field, by a coherent process such as the electron cyclotron maser (ECM) instability. 
This mechanism is known to be responsible for the radio emission at kHz and MHz frequencies 
from the magnetized planets (e.g., Jupiter) in our solar system. If the same mechanism is
responsible for radio bursts in ultra-cool dwarfs at frequencies of few GHz, the magnetic field
of a few kilo-Gauss is required on the dwarf surface \citep{hal06}, i.e., 
as high as those possessed by earlier-type classical M-type flare stars. 
There is also evidence that the radio pulses correlate with bright regions seen in the optical 
light curves \citep{har13,hal15}. The question of whether these bright spots are magnetic, 
as some models suggest \citep{kuz12}, remains to be answered. 
Also, processes to generate such strong large-scale magnetic fields in 
fully convective objects and to maintain them in largely neutral atmospheres are to be identified.

Among such bursty dwarfs, LSR J1835+3259 (hereafter LSR J1835) is one of the brightest 
in both near infrared and quiescent radio flux. Its spectral type M8.5 \citep{desp12} 
places it near the stellar-mass threshold and makes it a possible candidate for 
one of the warmest brown dwarfs. It demonstrates prominent periodic radio bursts, corresponding 
to a very rapid rotation with the period of only 2.84 hr \citep{hal15}.
Such a fast rotation (three times faster than Jupiter) may indicate the dwarf's young age and, 
therefore, the lower mass for the given spectral class.
No X-ray emission associated with the presence of a magnetically heated corona has been detected so far. 
We have selected LSR J1835 to test the hypothesis that such bursty brown dwarfs possess strong 
magnetospheres. 

The most reliable way to detect magnetic field on a cool dwarf is to measure its light polarization 
due to the Zeeman effect, which affects both atomic and molecular lines in its spectrum. 
In this paper, we present a quantitative analysis of a set of spectropolairmetric data 
for LSR J1835 taken at the Keck telescope \citep{har15}.
We report the first detection of a strong magnetic field on this brown dwarf. Our results 
provide direct evidence that strong transient optical and radio emission bursts on such objects 
are powered by the surface magnetic field of several kG which is associated with 
an active region producing both optical and radio emission.

The paper is structured as follows. In Section~\ref{sec:obs}, we briefly describe 
observations and their reduction and calibration procedures. 
In Section~\ref{sec:anal}, we analyse spectral feature evolution and 
evaluate the significance of polarimetric signatures in the Na I 819 nm lines,
which provide evidence for the surface magnetic field detection. 
In Section~\ref{sec:mod_bdw}, we determine atmospheric parameters of LSR J1835 from 
the Na I 819 nm Stokes $I$ line profiles. This helps us to deduce the dwarf's mass and age 
from evolutionary tracks and establish that LSR J1835 is a young brown dwarf. 
In Section~\ref{sec:mod_em}, we present emission region maps obtained by inversions 
of optical emission line profiles. The spatial distribution of the emission indicates 
the presence of emission loops rooted in the surface magnetic region 
and extended above the surface. 
In Section~\ref{sec:mod_mag}, we model the Na I 819 nm Stokes $I$ and $V$ line profiles
taking into account the Paschen---Back effect and deduce a 5\,kG surface magnetic field 
on LSR J1835. In Section~\ref{sec:dis}, we discuss the significance of our results in the
context of magnetic activity in low-mass stars and possible magnetic interactions
of the dwarf with its environment (disk, planets). Finally,
in Section~\ref{sec:sum}, we summarize our results and conclusions.

\section{Observations}\label{sec:obs}

We carried out spectropolarimetric measurements during 6\,hr on two consecutive nights 
(3\,hr/night) on 2012 August 22 and 23 with the Low Resolution Imaging Spectropolarimeter (LRISp) 
at the 10 m Keck telescope, Mauna Kea, Hawaii. Measurements were made first at four angles 
of the half-wave plate to obtain linear polarization described by Stokes $Q/I$ and $U/I$ and 
then at two angles of the quarter-wave plate to obtain circular polarization described by Stokes $V/I$.
The individual exposure time was 10 minutes for each position of the wave plates.
Thus, we obtained 36 Stokes $I$ flux spectra and six measurements for each of the polarized 
Stokes $Q/I$, $U/I$, and $V/I$ during two complete rotation periods of LSR J1835 separated by about 21\,hr, 
or about seven rotational periods. The measurements were made simultaneously in the blue 
(380--776\,nm) and red (789--1026\,nm) arms of LRISp, with the spectral resolution 
of 0.6\,nm and 0.3\,nm, respectively. 

The data and their reduction and calibration procedures were presented and discussed 
in detail by \citet{har15}. The reduction procedure included flat-fielding, spectral order
extraction, wavelength calibration using standard spectrum lamps, removing cosmic-ray hits, 
removing spectral ripples due to optical interference using standard star spectra and Fourier analysis,
polarimteric calibration, and Stokes parameter calculation. Since the LRISp is a dual-beam polarimeter,
combining two-beam Stokes measurements taken at different angles of the wave plates significantly
reduces systematic errors. Our thorough reduction procedure has allowed us to achieve
the noise standard deviation for both linear and circular polarization of about 
$1\times10^{-3}$ (0.1\%) per pixel. 

Our data set includes observations of LSR J1835 and several standard stars \citep{har15},
including the active M3.5 dwarf EV Lac. The EV Lac polarized spectra are analyzed here
to demonstrate that our reduction procedure is adequate to detect a few kG field 
such as that observed on EV Lac \citep{jkv96}.

\section{Data Analysis}\label{sec:anal}

Because of the target's rapid rotation and low brightness, 
circular and linear polarization could not be measured (quasi-)simultaneously. Therefore, 
they were analyzed separately. Here we present an analysis of observed flux variability of 
several optical emission lines (H$\alpha$, H$\beta$, H$\gamma$, and Na I D$_1$ and D$_2$ lines)
in the blue-arm spectra and polarization in the near-infrared (NIR) neutral sodium 
lines Na I at 819--820 nm in the red-arm spectra. Several bands of the TiO, CrH, and FeH 
molecules were also observed in the NIR at 850--1000 nm. Modeling their polarization using 
previously developed techniques \citep{berd03,berd05,afr07,afr08,kuzber13} provides additional 
important constraints on magnetic region properties, 
which is a subject of a separate paper \citep{kuzm16}.

\begin{figure*}
\centering
\includegraphics[width=7.5cm]{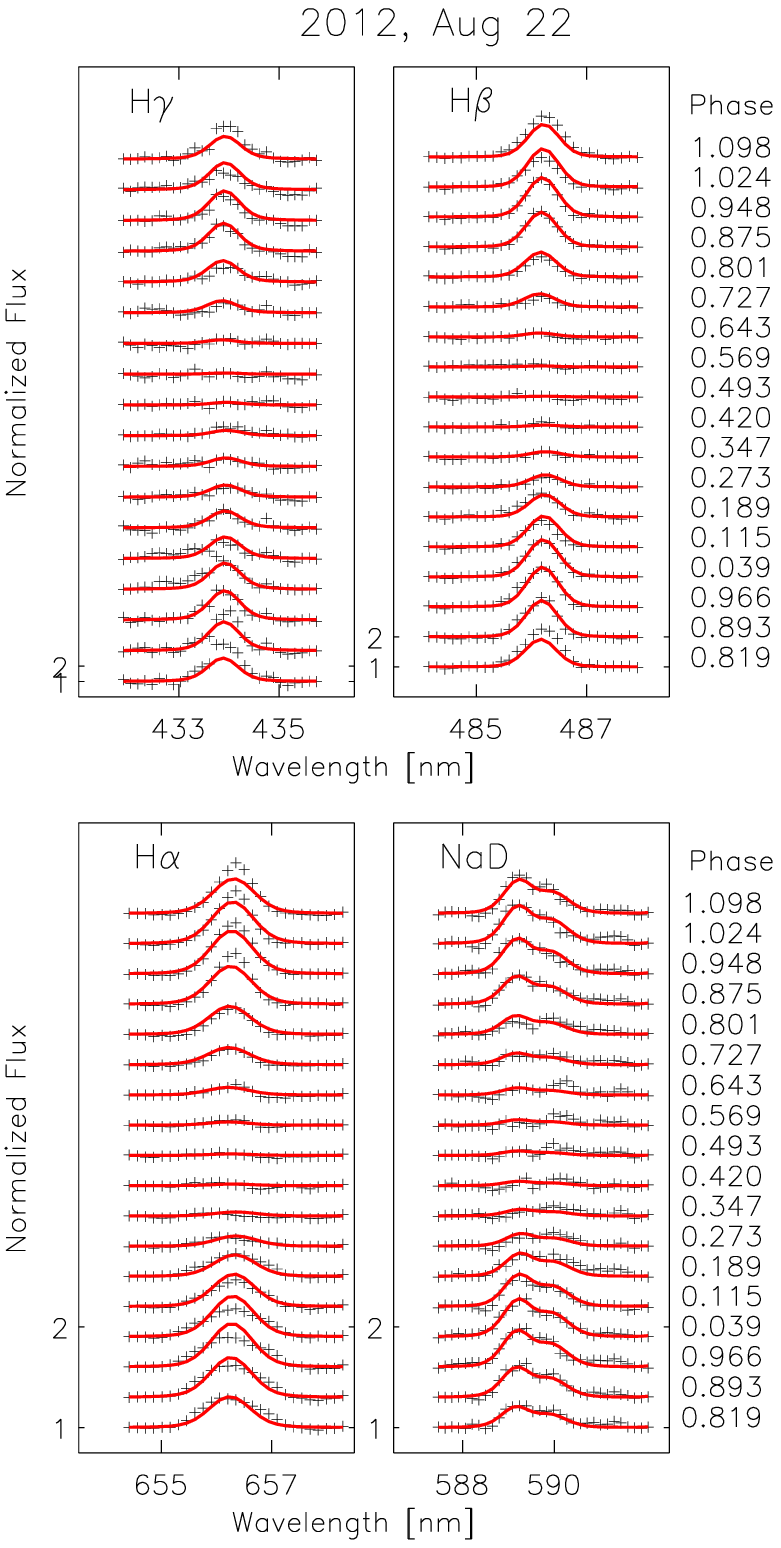}
\includegraphics[width=7.5cm]{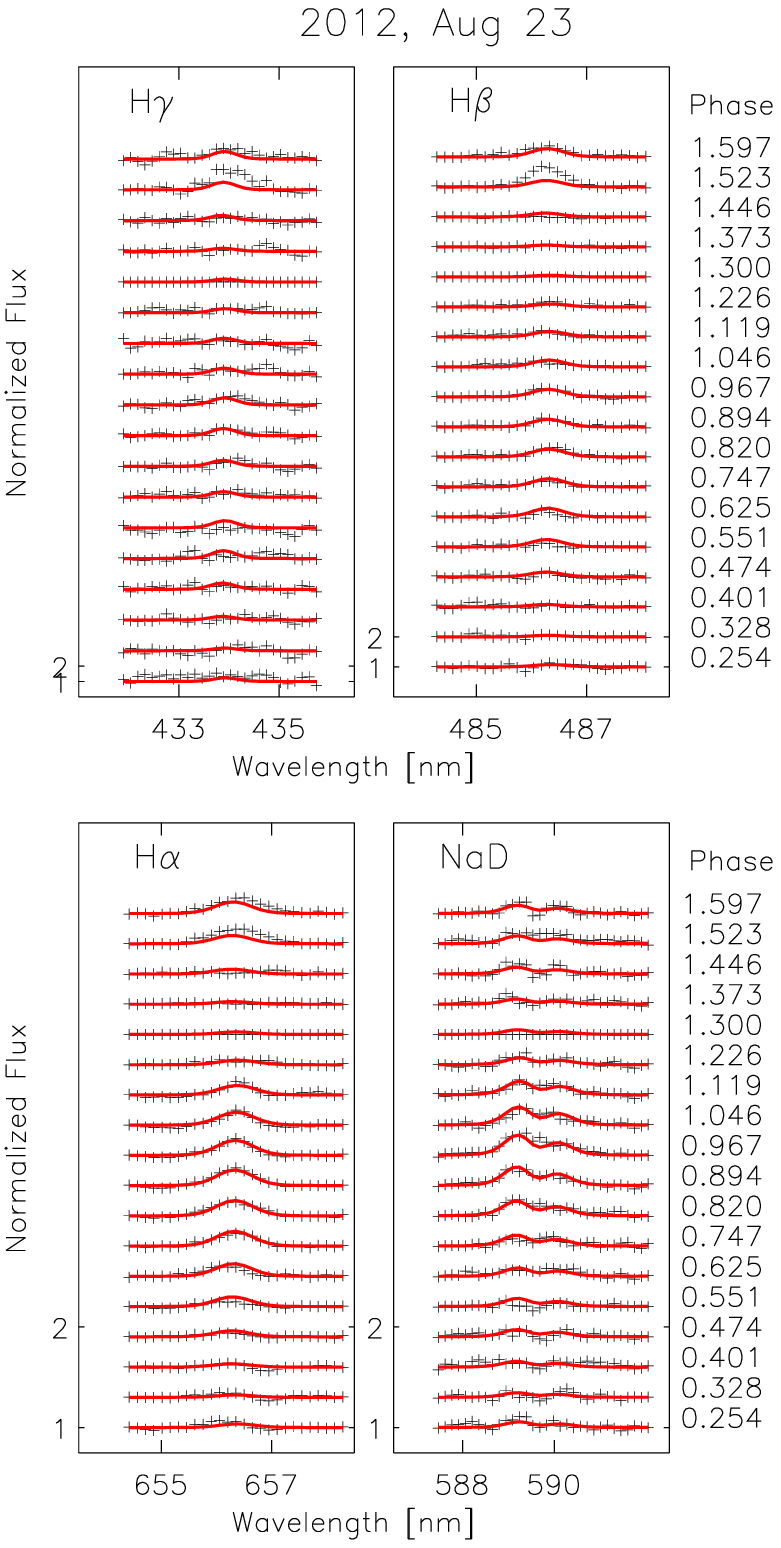}
\caption{
Non-thermal optical emission profiles observed on LSR J1835 
during two nights 2012 Aug 22 and 23 (on the left and right, respectively). 
The line profiles are obtained by subtracting the quietest state spectrum 
observed at the rotational phase 1.300 on 2012, Aug 23, to remove photospheric lines.
Crosses are observations, solid lines are best-fit profiles obtained from inversions
as described in Section~\ref{sec:mod_em}.
The maximum emission was observed on 2012 Aug 22, near the phase 0.0.
At nearly the same time, the magnetic field signal in the NIR Na I lines Stokes $V/I$
profiles was detected (phases 0.115-0.189 and 1.024-1.098, on 2012 Aug 22, see Fig.~\ref{fig:na_prof}).
On the second night, the emission of the primary active region (near phase 0.0) 
has reduced, and a new emission region has emerged at the opposite side of the dwarf 
(near phase 0.6).
}
\label{fig:em_prof}
\end{figure*}

On the first night, 2012 August 22, the optical emission observed in the blue arm
have indicated the presence of a ``hot spot'' (an emission region) centered at the same 
rotational phase ($\sim$0.0) as observed one month earlier by \citet{hal15} 
in both optical and radio. We used their ephemeris to calculate rotational phases 
for our observations. On the following night, 2012 August 23, 
the emission near the phase 0.0 significantly declined, while near the phase 0.6 
(opposite hemisphere) a rise of the emission was seen during the last two exposures. 
To analyze the emission variability in the blue-arm spectra, we have subtracted 
the most quiescent state spectrum, observed at the phase of 1.300 during the second night, 
from the 36 Stokes $I$ flux spectra. The resulting emission profiles are shown 
in Fig.~\ref{fig:em_prof}. They have been used to recover
emission maps as described in Section~\ref{sec:mod_em}.

\begin{figure}
\centering
\includegraphics[width=8cm]{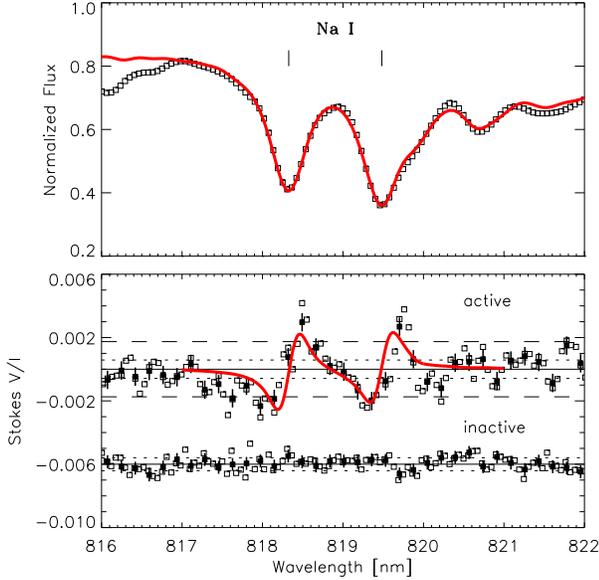}
\caption{
The NIR Na I line profiles of LSR J1835: observed (symbols) and modeled (thick red lines). 
Thin symbols are unbinned data, and thick symbols with error bars are binned by a factor of three.
The model profiles are shown for $T_{\rm eff}$=2800\,K, $B$=5.1\,kG, and $f$=0.11. 
The observed normalized flux spectrum (top panel) is averaged over all measured Stokes $I$ spectra
(SNR$\>$6000).
In the bottom panel, average Stokes $V/I$ profiles are shown.
The top Stokes $V/I$ spectrum is an average of two measurements near the phase 0.0 obtained 
on 2012 Aug 22, corresponding to the active state of the dwarf (see Fig.~\ref{fig:em_prof}). 
The bottom observed Stokes $V/I$ spectrum is an average of four measurements at other phases,
corresponding to an inactive state of the dwarf.
Horizontal solid lines indicate zero polarization, while dotted and dashed 
lines are the $\pm 1\sigma$ and $\pm 3\sigma$, respectively, noise levels per pixel.
}
\label{fig:na_prof}
\end{figure}

\begin{figure}
\centering
\includegraphics[width=8cm]{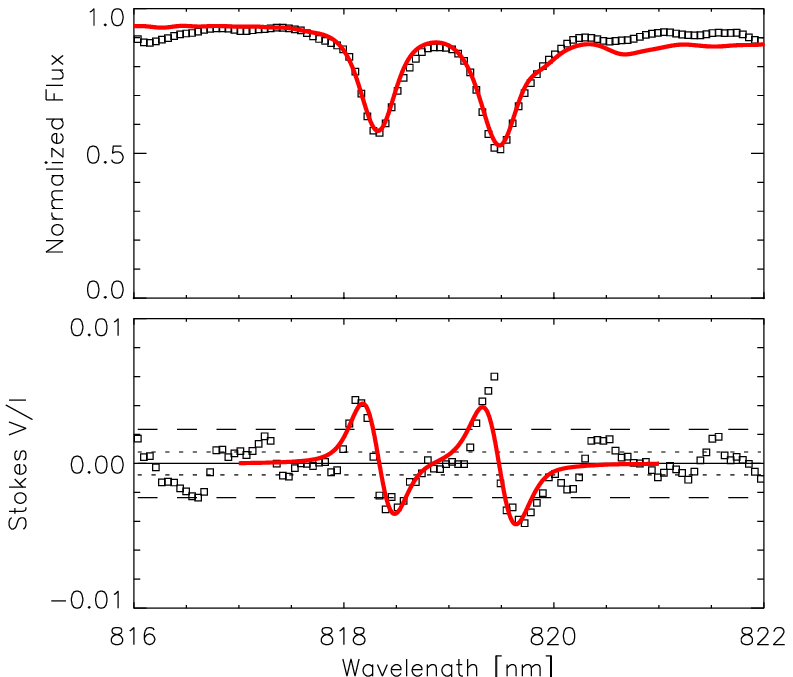}
\caption{
Same as Fig.~\ref{fig:na_prof} for EV Lac. The model Stokes profiles are shown for 
$T_{\rm eff}$=3300\,K, $B$=3.8\,kG, $\gamma=0^\circ$, and $f$=0.28.
}
\label{fig:evlac}
\end{figure}

The maxima of the blue emission near the phase 0.0 during the first night 
were accompanied by clear Zeeman effect signatures in circular polarization 
of the NIR Na I lines at 819--820 nm (Fig.~\ref{fig:na_prof}), which appears as a doublet. 
These are the strongest (and the only relatively unblended) atomic features 
in the red part of the observed polarized spectrum. In particular, significant 
Stokes $V/I$ profiles were detected in both lines 
during the active state at the two phases 0.11--0.19 
(in the beginning of the night) and 1.02--1.10 (3\,hr later) on August 22. 
The lower activity state Stokes $V/I$ averaged over the remaining four circular polarization 
measurements showed no signal above the 1$\sigma$ noise level.
The Stokes $Q/I$ and $U/I$ near the active state may have had some signals too, 
but their significance in the individual exposures is low. When averaged over the six 
corresponding measurements, no significant polarization was found in these profiles. 
This could also be in part because of cancellation of the opposite polarity signals 
in different rotational phases. The average Stokes $V/I$ profiles are shown 
in Fig.~\ref{fig:na_prof}.

Since the spectral PSF is oversampled \citep{har15}, 
binning the Stokes $V/I$ spectrum by 3 pixels reduces the noise per spectral bin down to 
$6\times10^{-4}$ (0.06\%) without a loss of the spectral resolution. 
The Stokes $V/I$ peak-to-valley average amplitude detected during the active state of 
LSR J1835 is $6.2\times10^{-3}$ (0.62\%), see Fig.~\ref{fig:na_prof}, 
so the detection significance is about five standard deviations for one Stokes $V/I$ lobe. 
Considering that we measure two profiles with four lobes, 
the overall significance of this detection is higher. 

On the second night, the optical emission near phase 0.0 has significantly decreased, 
and no magnetic signal in the NIR Na I lines was observed (within the noise level),
thus, indicating a possible decay of the magnetic region. 
During the last two exposures, the emission started to increase at the opposite
hemisphere (near phase 0.5), indicating a possible emergence of a new bursting region.
Yet the low hydrogen emission level was not accompanied by a detectable magnetic signal
in the NIR Na I lines. 

It is worth mentioning that the NIR Na I and optical Na I D$_1$ and D$_2$ share the same 
electronic doublet $P$-state
(with the orbital momentum $L=1$, spin $S=1/2$, and principal quantum number $n=3$), 
This is the upper state for the optical Na I lines and the lower state for the NIR Na I lines.
Therefore, we have also searched for emission and absorption variations in the NIR line. 
There were no variations found above the noise level. Thus, we conclude that 
the detected Stokes $V/I$ profiles are formed due to absorption in the magnetized photosphere. 
This is supported by the fact that the polarization sign observed in the NIR Na I lines
and in absorption CrH lines \citep{kuzm16} is the same.

To verify our calibration procedure we also measured polarization of a very active, 
flaring M3.5 red dwarf EV Lac during our observing run. 
This star is known to possess a magnetic field of 3.8$\pm$0.5 kG with at least 50$\pm$13 \%\ 
coverage of the stellar surface \citep{jkv96}.
We detected Stokes $V/I$ peak-to-valley amplitude of $9.6\times10^{-3}$ (0.96\%). 
No significant Stokes $Q/I$ or $U/I$ were detected. The noise per pixel achieved is 
$9\times10^{-4}$ (0.09\%). The Stokes profiles are shown in Fig.~\ref{fig:evlac}). 
Our measurement for EV Lac agrees with the result by \citet[][see Section~\ref{sec:mod_mag}]{jkv96}, 
assuming that field is predominantly radial near the center of the visible stellar disk.

\section{LSR J1835+3259 Atmosphere Parameters, Mass, and Age}\label{sec:mod_bdw}

To determine the atmosphere parameters of LSR J1835, we
compute Stokes profiles with the STOPRO code capable of solving 
a full set of polarized radiative transfer equations for both atomic and molecular lines
including magnetic field effects and under the LTE assumption  
\citep{frut00,berd03}. To evaluate possible systematic errors, 
we have used three grids of model atmospheres with the solar 
metallicity and other element abundances: 
the PHOENIX models AMES-Cond \citep{all01} with clear atmospheres and
BT-Settl with clouds \citep{all12} as well as MARCS clear models \citep{gus08}.

The NIR Na I lines are blended by numerous weak molecular lines. We have identified major
contributions from the TiO bands of the triplet $\gamma$ and singlet $\delta$ systems.
In particular, the strongest band heads were observed at 820 nm from the $\delta(1,0)$
band and at 820.5 nm from the $\gamma(2,1)$ band. Overall, we have included about
7000 lines from the laboratory list of \citet{dav86} and simulated list by \citet{plez98}.
Magnetic level splitting and transition probabilities were calculated using a perturbation
theory described by \citet{berd02} and \citet{berd05}. This theory is accurate to explain Stokes
profiles in the TiO lines observed with high resolution in sunspots \citep{berd00,berd06a}
and starspots \citep{berd06b}. 
We have also verified wavelengths of simulated lines by comparing the list with high-resolution
spectra of M dwarfs obtained with the ESPADONS spectropolarimeter at the CFHT \citep{berd06b}.
At the spectral resolution of our data, we do not detect
significant TiO polarization due to blending. However, their absorption influences
Stokes profile shapes of the Na I lines.

\begin{figure}
\centering
\includegraphics[width=8cm]{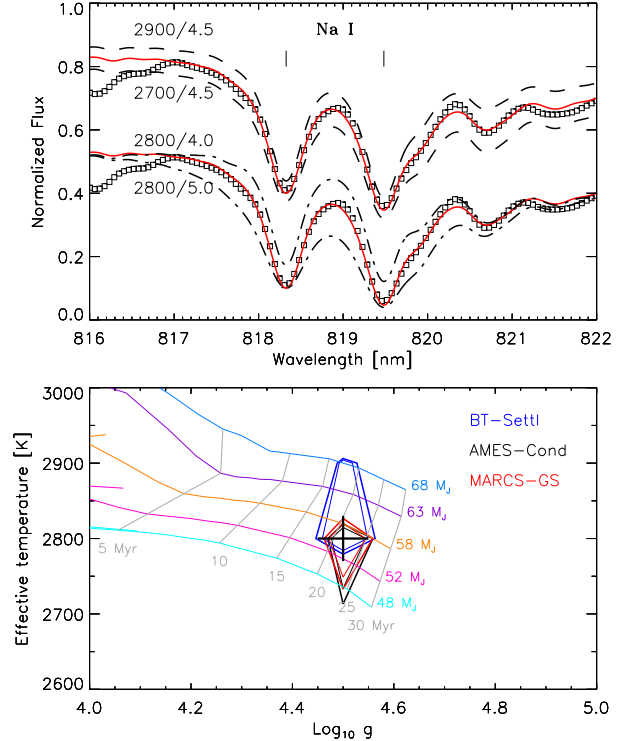}
\caption{
Top panel: the best-fit model spectra for $T_{\rm eff}$=2800\,K and $\log g$=4.5 
(solid red lines) compared to the observed Na I + TiO Stokes $I$ normalized flux line profiles 
(squares) and the model spectra for 
$T_{\rm eff}$=2700\,K and 2900\,K (dashed lines) and $\log g$=4.0 $\log g$=5.0 (dashed-dotted lines)
using AMES-Cond models. Lower spectra are shifted down by 0.3 for clarity.
Bottom panel: the confidence contours for the three model grids AMES-Cond (black), BT-Settl (blue),
and MARCS-SG (red) in the $\chi^2$-landscape. The 67\%\ level (1$\sigma$) is the thin inner contour, 
and the 99\%\ level (3$\sigma$) is the thick outer contour.  
The minima for the three model types overlap at the best-fit solution and define its uncertainty
(marked with the cross).
Contours for the AMES-Cond and MARCS-SG models are almost identical. 
In addition, evolutionary tracks by \citet{bar17} for young brown dwarfs with different masses
(thin solid curves of various colors) and corresponding isochrones (light gray solid lines) are shown and labeled. 
The atmospheric parameters of LSR J1835 correspond to a brown dwarf 
with the mass of 55$\pm$4\,$M_{\rm J}$ and age of 22$\pm$4\,Myr.
}
\label{fig:na_prof_chi}
\end{figure}

By simultaneously modeling the Na I and TiO lines in the
816--822 nm regions, we have found that this region is
extremely sensitive to both the effective temperature $T_{\rm eff}$
and gravitational acceleration $\log g$. In particular, the Na I line wings 
are mostly sensitive to the gravity, while the weak TiO bands
are sensitive to the temperature (Fig.~\ref{fig:na_prof_chi}, upper panel). 
These features form in deep atmospheric layers,
so they reliably probe the gravity and temperature near optical depth 1, i.e. in the photosphere.
Earlier, the sensitivity of these lines to the gravity 
and age was pointed out by \citet{luh12} based on a comparison of line profiles 
for late-M young cluster members and field dwarfs. We have searched 
for the best fit to the observed Stokes $I$ normalized flux profiles
within each of the three model grids by means of minimization of the $\chi^2$ 
functional, according to \citet{press}.
 
All the model grids have shown well-constrained $\chi^2$ minima for the model 
with $T_{\rm eff}$=2800\,K and $\log g$=4.5 (Fig.~\ref{fig:na_prof_chi}, lower panel).
The confidence levels for the three model grids nicely overlap around the minimum. 
For the clear atmospheres, such as the AMES-Cond and MARCS-SG models, contours are almost identical.
This allows us to narrow down the uncertainties to $\pm$30\,K for $T_{\rm eff}$ 
and $\pm$0.05 for $\log g$. We note that non-LTE effects were not accounted for
in our modeling, but they may not influence our result, 
because we have achieved good fit to both weak molecular lines and strong Na I lines.
For EV Lac, we found the best fit for the model with $T_{\rm eff}$=3300\,K and $\log g$=4.5,
i.e., what can be expected for an M3.5 dwarf with cool spots.

We note that strong molecular bands, such as TiO, CrH, and FeH bands in the
830--1000\,nm region (included in our data), are also sensitive atmospheric diagnostics.
However, they contain a significant contribution from cool magnetic starspots
that increases the number of unknowns and their errors. 
In addition to the temperature sensitivity, metal hydrides are known to be 
quite sensitive to the gravity acceleration \citep[e.g.,][]{bell85,schi97},
and they are also highly magnetically sensitive \citep{berd05,afr08,kuzber13}.
Our analysis of the observed Stokes $I$ in these molecular bands results in 
the temperature of the non-magnetic photosphere $T_{\rm phot}=2900 +150/-250$\,K,
magnetic spot temperature $T_{\rm spot}=2200 +250/-150$\,K,
spot filling factor $f_{\rm spot}=0.6$, 
and $\log g$ of 4.5--5.0, equally good within the errors \citep{kuzm16}. 
The $T_{\rm eff}$=2800\,K and $\log g$=4.5 determined in the present paper are more accurate,
because they are based on weak, higher-excitation TiO bands 
and wings of the subordinate atomic lines (also higher-excitation lines),
which form deeply in the atmosphere and are less sensitive to starspots. 

Knowing the atmosphere parameters, we can now deduce the age and mass of LSR J1835 from
evolutionary models. Different models, e.g., by
\citet{bur97} and \citet{bar17}, indicate a young age of the dwarf.
The models by \citet{bar17} are based on self-consistent calculations, which couple 
numerical hydrodynamics simulations of collapsing pre-stellar cores
and stellar evolution models of accreting low-mass stars and brown dwarfs 
with the age up to 30 Myr. Models with two accretion scenarios were considered: 
a cold accretion model, with all accretion energy is radiated
away, and a hybrid accretion, with the amount of accreted energy depending 
on the accretion rate. In particular, the hybrid scenario helps explain
observed luminosity spread in young clusters and FU Ori eruptions.
Using these models, we determine that
an object with $T_{\rm eff}$=2800$\pm$30\,K and $\log g$=4.50$\pm$0.05 
is a young brown dwarf with the mass, radius, and age as follows.

\begin{center}
$M=55\pm 4 M_{\rm J}$,\\ 
$R=2.1 \pm 0.1 R_{\rm J}$,\\ 
$t=22 \pm 4 {\rm Myr}$.\\
\end{center}

Relevant evolutionary tracks and isochrones for the hybrid accretion scenario 
are shown in the lower panel of Fig.~\ref{fig:na_prof_chi}, but the differences
between the cold and hybrid scenario tracks are significantly smaller than our
error bars. 
For comparison, the older models by \citet{bur97} predict a similar age (27 Myr) 
but a somewhat smaller mass (44 $M_{\rm J}$) and radius (1.9 $R_{\rm J}$).
Therefore, we conclude
that LSR J1835 is a young brown dwarf in the end of accretion phase
(see also Section~\ref{sec:dis}).

We emphasize that LSR J1835 atmosphere parameters
significantly differ from those typical for late stellar-mass M dwarfs. 
For instance, a red dwarf star with $T_{\rm eff}$=2800\,K
would have the spectral class M5 and $\log g$=5.0, 
while an M8 star with the mass 0.08\,$M_\odot$ would have
$T_{\rm eff}$=2400\,K and $\log g$=5.25. 
As concerns the kinematics, UVW velocities of 
brown dwarfs in the solar neighborhood are rather disperse \citep{zap07},
and LSR J1835 falls in the middle of the distribution.

\begin{figure*}
\centering
\includegraphics[width=8cm]{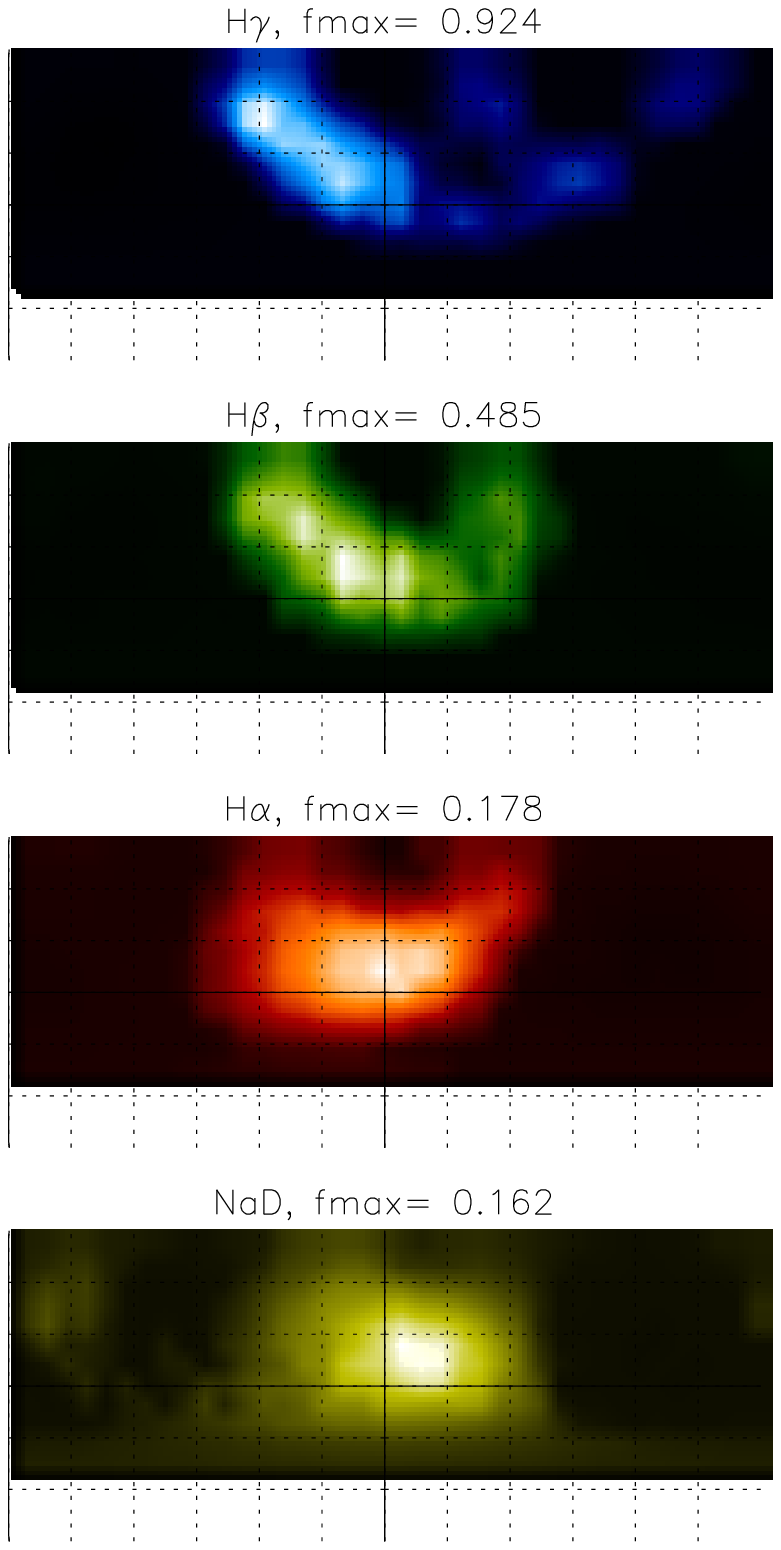}
\includegraphics[width=8cm]{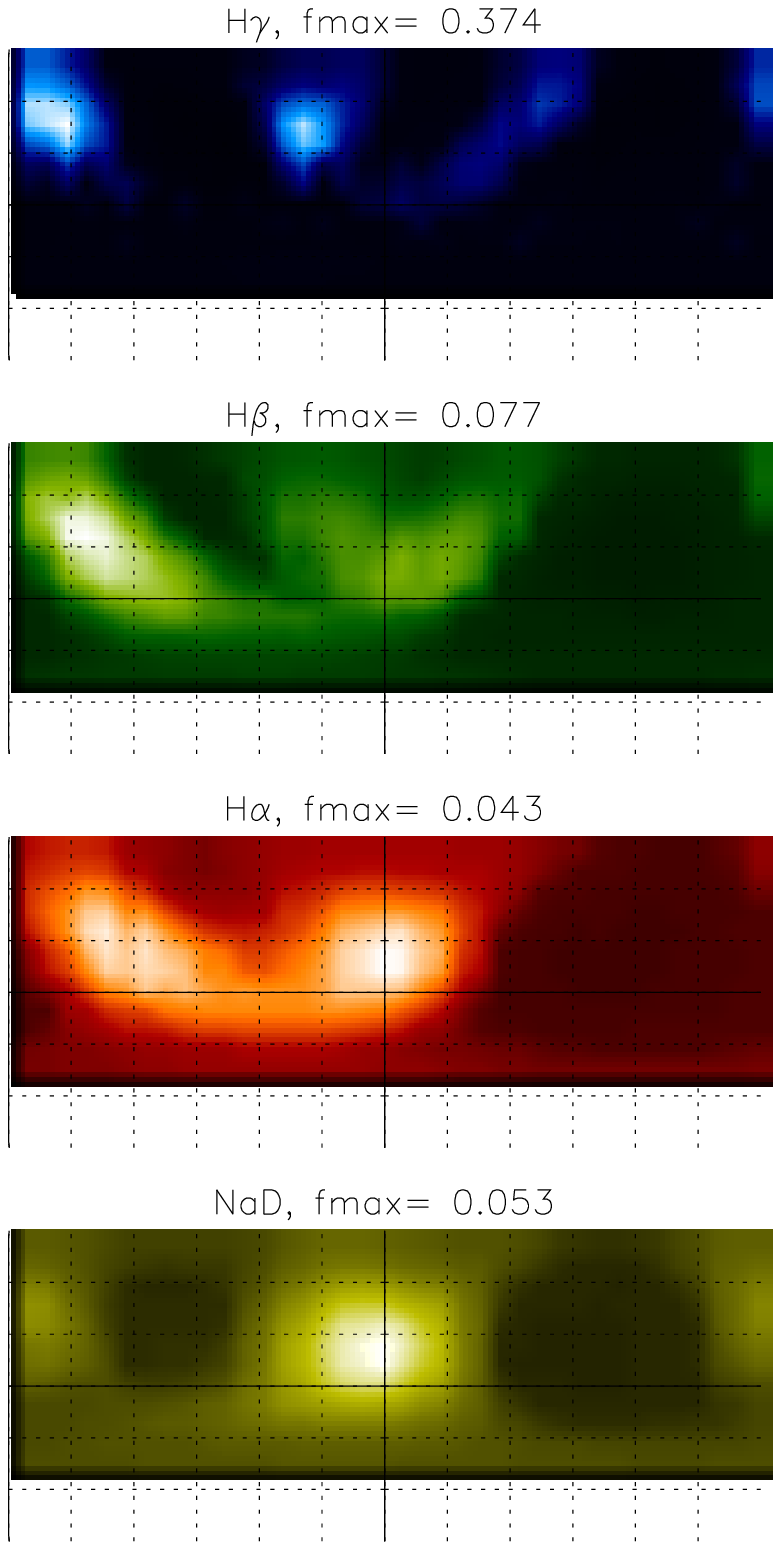}
\caption{
{Non-thermal optical emission maps recovered from the profiles shown in Fig.~\ref{fig:em_prof} 
during two nights 2012 Aug 22 and 23 (on the left and right, respectively).
Maps are plotted in a cylindrical projection: the coordinate grid is shown with 
the $30^\circ$ interval in both latitude and longitude (dotted lines). The phase=0.0
meridian and the equator are shown with solid lines. The inclination of the rotational axis 
is $50^\circ$. 
The maximum filling factor of the hot flare-like plasma (see Section~\ref{sec:mod_em})
is provided above each image and corresponds to the lightest area in the images. 
The minimum filling factor (0.0) corresponds to the darkest areas in the images.}
}
\label{fig:em_maps}
\end{figure*}

\begin{figure}
\centering
\includegraphics[width=8cm]{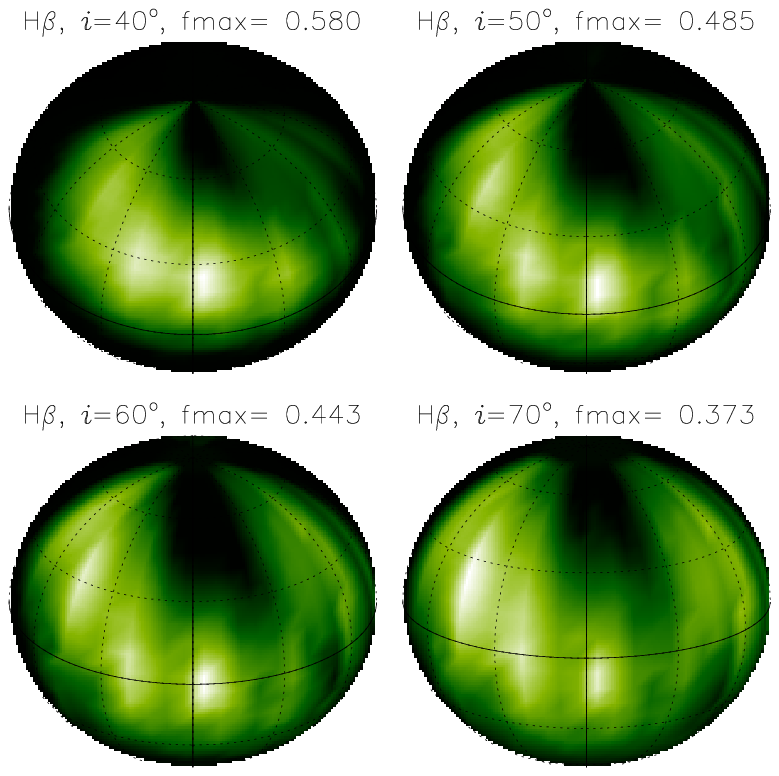}
\caption{
H$\beta$ emission maps as in Fig.~\ref{fig:em_maps} for different inclination angles
of the dwarf's rotational axis, for the night 2012 Aug 22. 
The images are shown at the phase 0.0 in an orthographic projection, i.e.,
this is how an observer would see the dwarf's surface for a given inclination angle.
The dwarf is assumed to rotate clock-wise as seen from the visible pole.  
}
\label{fig:em_inc}
\end{figure}

\section{Emission Maps and Loops}\label{sec:mod_em}

The two time series of the emission line profiles obtained during two complete 
rotations of the dwarf (Fig.\ref{fig:em_prof}) are suitable for recovering the topology 
of the emitting regions on LSR J1835 using an indirect Doppler imaging technique. 
Here we employ the inversion method based on Occam's razor principle
by \citet{berd98}, which was successfully employed for indirect imaging of cool active
stars using spectroscopic and photometric measurements \citep[e.g.,][]{berdliv05}. 

Inversion techniques, when applied to spectra, rely on the knowledge of local line profiles 
depending on physical conditions (e.g., temperature, density, magnetic field, etc.) 
at a given location on the surface. 
Such profiles have been recently computed using the most advanced 
NLTE radiative-hydrodynamic model of a typical M-dwarf flare by \citet{kow15}. 
They have demonstrated that the F13 flare model based on precipitating non-thermal 
electrons with the energy flux of 10$^{13}$ erg cm$^{-2}$ s$^{-1}$ successfully 
reproduces the color temperature and Balmer jump ratio observed in M-dwarf highly 
energetic flares. 

We have found that this model is also suitable for interpreting 
emission profiles of LSR J1835. In particular, we consider model line profiles 
at rising, peak, and decaying stages of the flare (RF, PF, and DF, respectively). 
These profiles differ by equivalent widths and broadening in the line cores 
(Doppler broadening) and in the line wings (Lorentz broadening) due to evolution 
of the plasma temperature and density during the flare. 

Here we summarize most relevant characteristics of the flare stages and refer to 
other details in \cite{kow15}.
The model predicts that most of the non-thermal electron energy is first deposited into 
hydrogen ionization and then into raising the temperature, up to 180,000\,K, 
resulting in massive ionization of hydrogen and helium (and other elements) and 
a sequence of explosive events during the rising stage of the flare (models RF, F13 0.2s and 1.2s).
At the peak of the flare (model PF, F13 2.2s)
the temperature settles at 12,500\,K and 13,000\,K, and then eventually drops to
7000\,K at the decaying stage of the flare (model DF, F13 4s).
Correspondingly, the continuum and line emission 
are observed due to recombinations and radiative decays during the flare.
As the flare decays, the emission decreases and the line profiles become narrower. 
Note that the time scale of the model flare (0.2s to 4s after the flare peak) 
is too short as compared to flare
durations in M dwarfs and the brown dwarf discussed here. This can be overcome by
integrating over a series of flaring events triggered by an initial reconnection 
\citep{kow15}. Nevertheless, this model provides us with realistic emission line profiles 
at different flare stages suitable for our analysis.

We have carried out inversions of the time series line profiles for each night 
and each emission line separately, thus obtaining four maps corresponding to 
different heights in the dwarf atmosphere (according to the response of the
atmosphere to the flare) for each full rotation. 
They are presented in Fig.~\ref{fig:em_maps}. A relatively low-resolution
map coordinate grid of $6^\circ\times6^\circ$ was used to reduce the number 
of independent parameters.
The emission flux from each map pixel was calculated using one selected model line profile
for a given emission line weighted by a filling factor $f$. Thus, the recovered image
is a map of the filling factor for a given model emission line. 
This approach allows us to identify the activity stage of the emission bursts observed 
on LSR J1835 and evaluate the temperature and density needed to produce such emission.

Out of all the F13 model line profiles, the decaying flare stage profiles (DF, F13 4s)
were found on average to provide the best fits to the observed profiles.
The PF and RF profiles were found to be too broad (plasma is too hot and too dense) 
to obtain satisfying fits to the observed profiles. Finer details in the emission evolution 
were found to deviate also from the best DF F13 4s model:
in the beginning of observations the measured profiles are somewhat broader
than the model (indicating temperature $>$7000\,K and higher density), 
while toward the end of observations the measured profiles are narrower 
(indicating temperature $<$7000\,K and lower density).
We found that by adjusting Lorentz broadening we can achieve better fits.
This is simply because a brown dwarf atmosphere is more compact than an early M-dwarf
atmosphere, leading to a smaller scale height and larger density gradient.
Furthermore, temperature and density are also expected to evolve as the emission 
slowly decays. Most recently (after our analysis was completed) 
the F13 model was further updated by \citep{kow17} with a new
electric pressure broadening prescription allowing for broader emission lines.
This theoretical improvement is in agreement with our empirical adjustments.

The dwarf's projected rotational velocity $v\sin i$ of $50\pm5$\,km\,s$^{-1}$ has 
been estimated from fits to the NIR Na I flux profiles (Section~\ref{sec:mod_mag}).
With this rotational velocity and the spectral resolution of 0.5--0.6\,nm in the
blue arm, there is no measurable Doppler shifts to resolve emission latitudes. 
However, since the dwarf rotation is much faster than the emission decay,
the emission amplitude rotational modulation and its visibility during a range
of phases provide a good constraint on its longitude and somewhat limited
constraint on the latitude (in fact, they are projected latitude and longitude,
since emission is formed above the photosphere). The longitudinal resolution 
is about $10^\circ$ due to the 10\,minute sampling of the profiles.

The inclination of the rotational axis to the line of sight is unknown.
To investigate the sensitivity of the result to this parameter, we obtained
maps for the inclinations of $40^\circ$, $50^\circ$, $60^\circ$, and $70^\circ$.
Maps for the H$\beta$ emission at different inclination angles are presented 
in Fig.~\ref{fig:em_inc} as examples.
The variations of the active region geometry depending on the inclination are
typical for Doppler imaging \citep[see][]{berd98}, i.e., the emission latitudes
and maximum filling factors adjust themselves accordingly to maintain the same emission 
flux at a given rotational phase. Hence, the emission region area increases with the 
inclination angle, while the maximum filling factor decreases. 
Since the emission is observed at a wide range of rotational phases, at lower inclination
its area is compact and shifted to higher latitudes, while at higher inclinations 
the emission region is broad and shifted to lower latitudes.
However, the relative distribution of the emission in different spectral lines remains 
about the same, which is an important constraint on the topology of the emitting region.

The recovered emission maps indicate the presence of dense and hot plasma, 
of at least 7000\,K, which is much hotter than the surrounding atmosphere. 
The relative latitude of the region near the phase 0.0 varies depending on the
line: emission of the higher-excitation lines (H$\gamma$ and H$\beta$)
occurs at higher latitudes and occupy a larger area, while emission of 
the low-excitation lines (Na I D$_1$ and D$_2$) occupies a compact region
at lower latitudes. The emission in H$\alpha$ is intermediate to these two extremes
with the brightest spot close to that of the Na I D lines.

\begin{figure*}
\centering
\includegraphics[width=8cm]{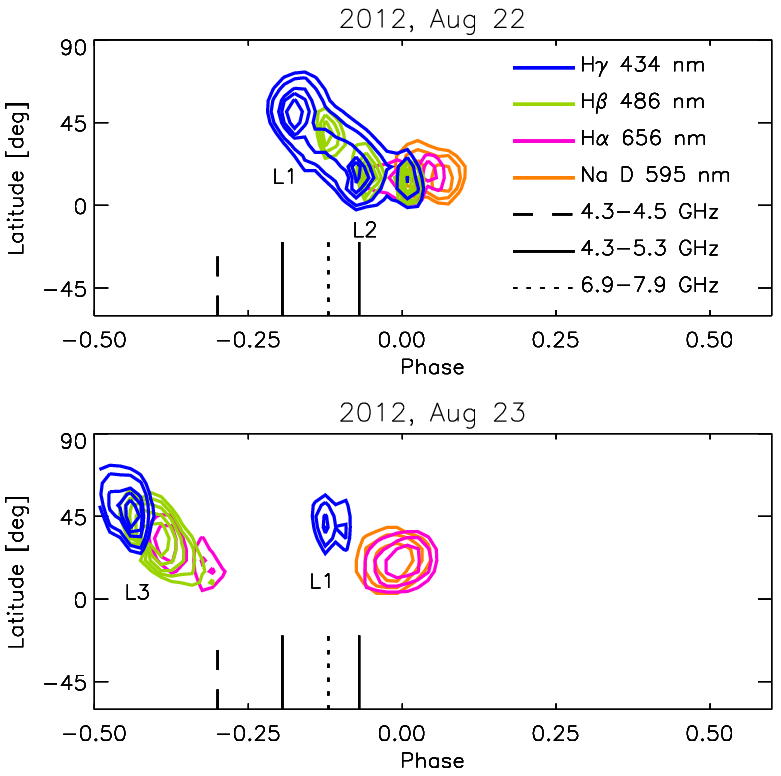}
\includegraphics[width=8cm]{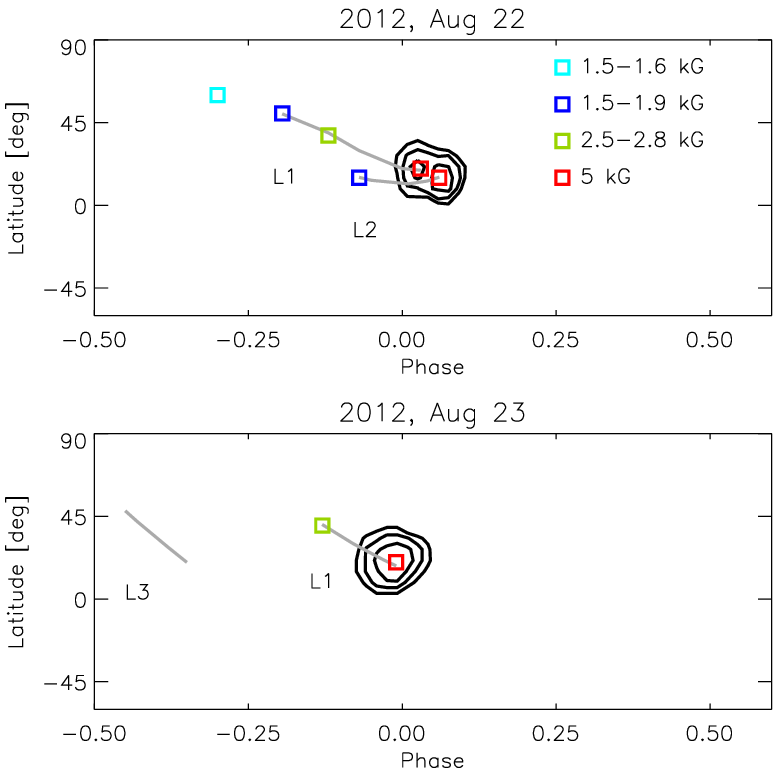}
\caption{
Left: emission contours around their maxima from the maps in 
Fig.~\ref{fig:em_maps} for the Na D, H$\alpha$, H$\beta$, and H$\gamma$ lines. 
There were possibly two emission loops L1 and L2 within the primary region near 
the phase 0.0 on 2012 Aug 22. On the following night, these loops significantly decayed,
and the new loop L3 emerged at the almost opposite side of the dwarf.
The phases at which radio bursts were observed a month earlier \citep{hal15} are marked
with vertical lines at the bottom of the plots. They coincide with the phases of the
H$\beta$ and H$\gamma$ emission hot spots (see discussion in the text).
Right: the magnetic spot map obtained from the NIR Na I line Stokes profiles
(black contours). 
The emission loops from the plots on the left are indicated as gray thick lines.
The magnetic field strength is indicated at different locations along the loops
as deduced from the the Na I Stokes profiles and the radio frequencies of the bursts.
}
\label{fig:em_cont}
\end{figure*}

We visualize the spatial connection of the emission in different lines by overplotting 
contours around their maxima in Fig.~\ref{fig:em_cont}. There are at least two emission 
loops (marked as L1 and L2) within the primary region near the phase 0.0 on the first 
observing night. The emission is apparently vertically stratified within the loops. 
These loops are seen under different angles at different rotational phases and
rotate with the dwarf in and out of view causing the emission bursts.
In the true, orthographic projections (Fig.~\ref{fig:em_inc}) the loops are
reminiscent of large filaments/prominences above the dwarf's surface,
as directly observed on the Sun. Such loops may extend one to three dwarf radii 
from the surface \citep{lyn15}.
The region near the phases 0.4--0.5 on the second night seems have
a vertical structure similar to that of the primary region. 

One month earlier, observations by \cite{hal15} detected emission activity near 
the phase 0.0. This provides a lower limit on the life time of the region.
For comparison, on the Sun, large prominences can last several months and are
often sources of large eruptions.
As described in Section ~\ref{sec:obs}, the phase of the emission region on LSR J1835 
coincides with the phase at which the Stokes $V/I$ was detected in the NIR Na I lines. 
Thus, the emission is clearly associated with a magnetic field. 
We model this magnetic signal in the following section.

\section{Magnetic Field}\label{sec:mod_mag}

\subsection{Paschen--Back Effect in the Na I lines}\label{sec:mod_pbe}

The NIR Na I line parameters (wavelength, lower level excitation energy, oscillator strengths, 
and electronic configurations) were taken from the Kurucz database \citep{kur93}.
They indicate that the upper doublet levels with $J'$=2.5 and 1.5 
(the fine structure levels) are split only by 0.05\,cm$^{-1}$. 
This implies that magnetic splitting of these levels will be comparable to the 
fine structure splitting at the field strength of 200\,G. Therefore, at stronger
fields the magnetic perturbation should be considered in the intermediate 
Paschen--Back regime (PBR).
The lower doublet levels with $J''$=1.5 and 0.5 are split by about 
17\,cm$^{-1}$, which requires fields stronger than 100\,kG for 
the PBR. Therefore, magnetic perturbations
of the lower levels can be safely described in the linear Zeeman regime (ZR).

\begin{figure*}
\centering
\includegraphics[width=4.5cm]{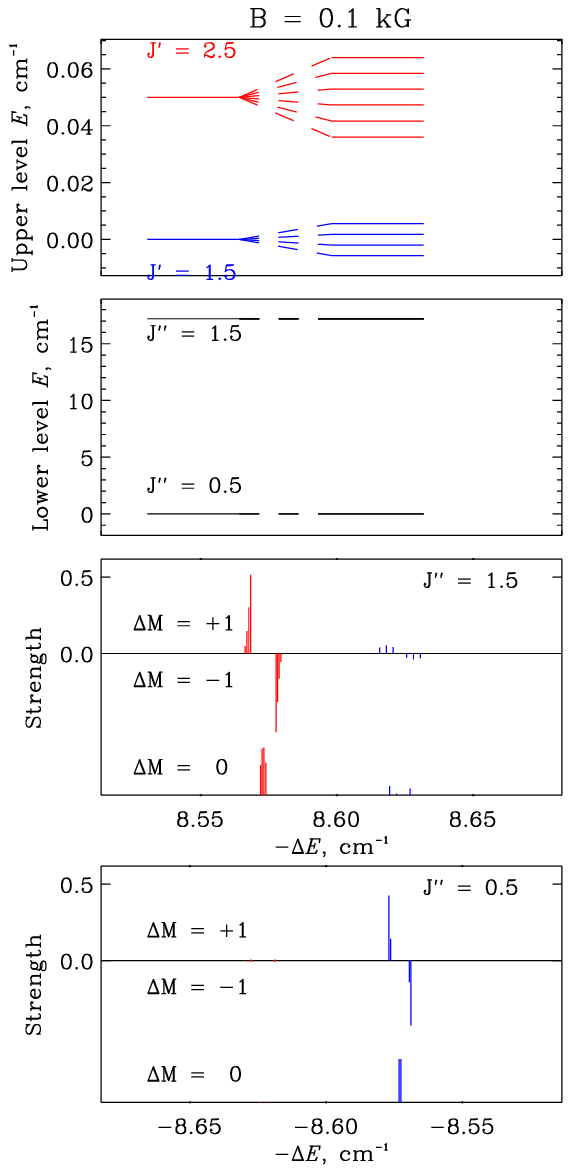}
\includegraphics[width=4.5cm]{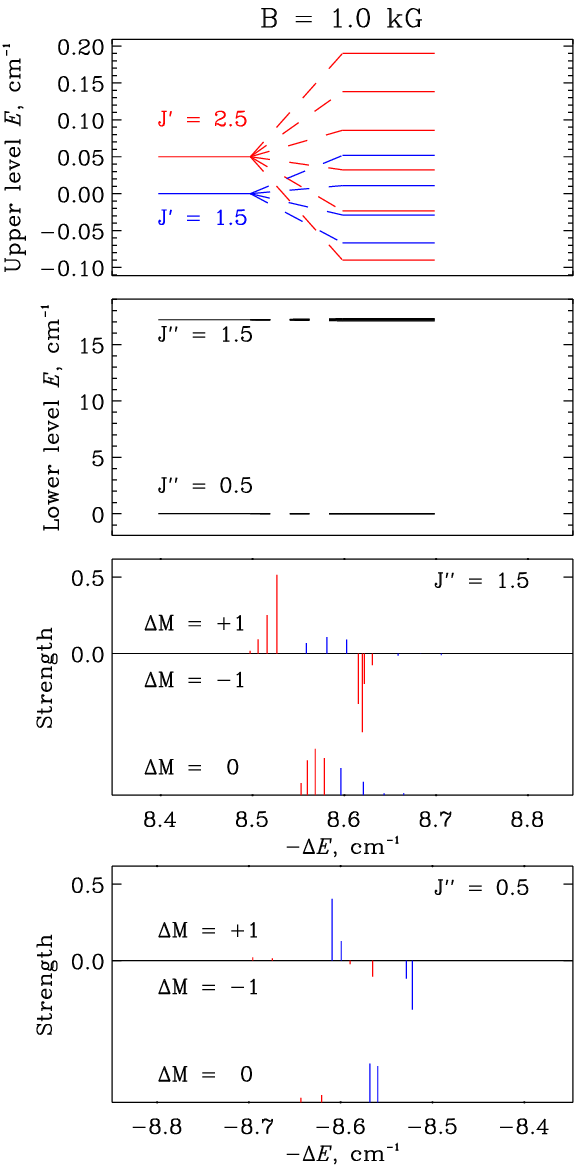}
\includegraphics[width=4.5cm]{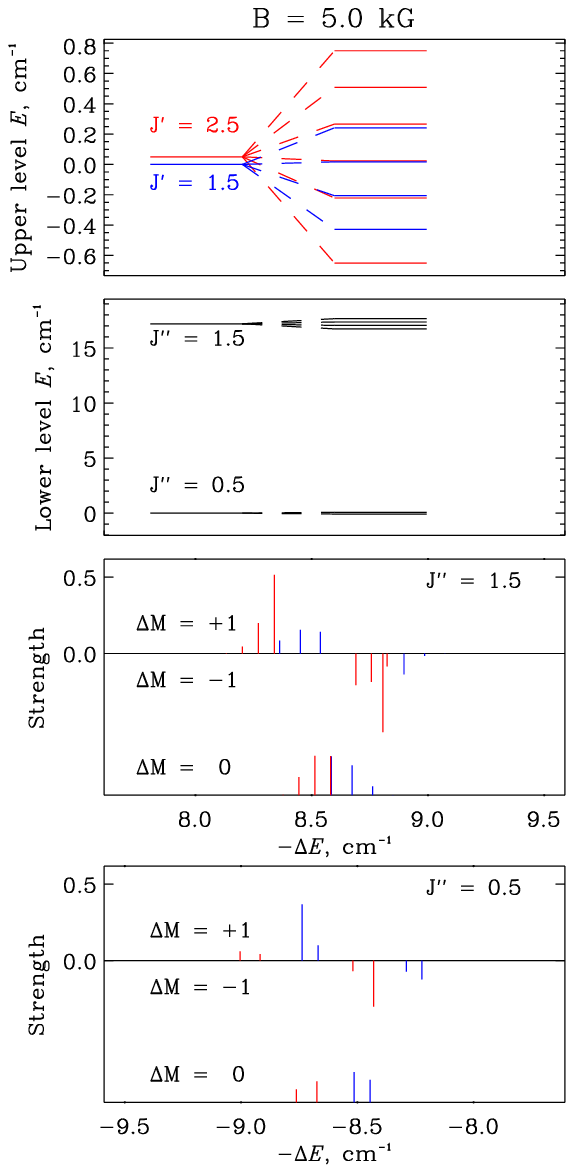}
\caption{
Magnetic splitting of the NIR Na I fine structure levels (top two rows) and
magnetic transitions (bottom two rows). The levels and corresponding transitions involving
magnetic sublevels of the $J'$=2.5 level are shown in red, and those of $J'$=1.5 are shown in
blue. The magnetic transitions for $\Delta M$=+1 are shown upwards, for $\Delta M$=--1
downwards, and for $\Delta M$=0 at the bottom of the plots.  
}
\label{fig:pbe}
\end{figure*}

It is easy to solve the intermediate PBR problem for doublet atomic levels, 
as described in textbooks \citep[e.g.,][]{sob92}. 
We show a few examples of level splitting of the NIR Na I lines in Fig.~\ref{fig:pbe}.
Several interesting effects in the PBR are worth to noting:
(1) mixing of the fine structure levels (i.e., $J'$ is no longer a good quantum number),
(2) mixing of magnetic components of different lines,  
(3) change of permitted line strengths as compared to the zero-field values, and
(4) appearance of forbidden transitions (here, with $\Delta J=J'-J''=2$).
As a result, line splitting undergoes complex nonlinear transformations
as the field strength increases. Therefore, taking into account these effects
is important for adequate interpretation of observed Stokes profiles.

\begin{figure}
\centering
\includegraphics[width=8cm]{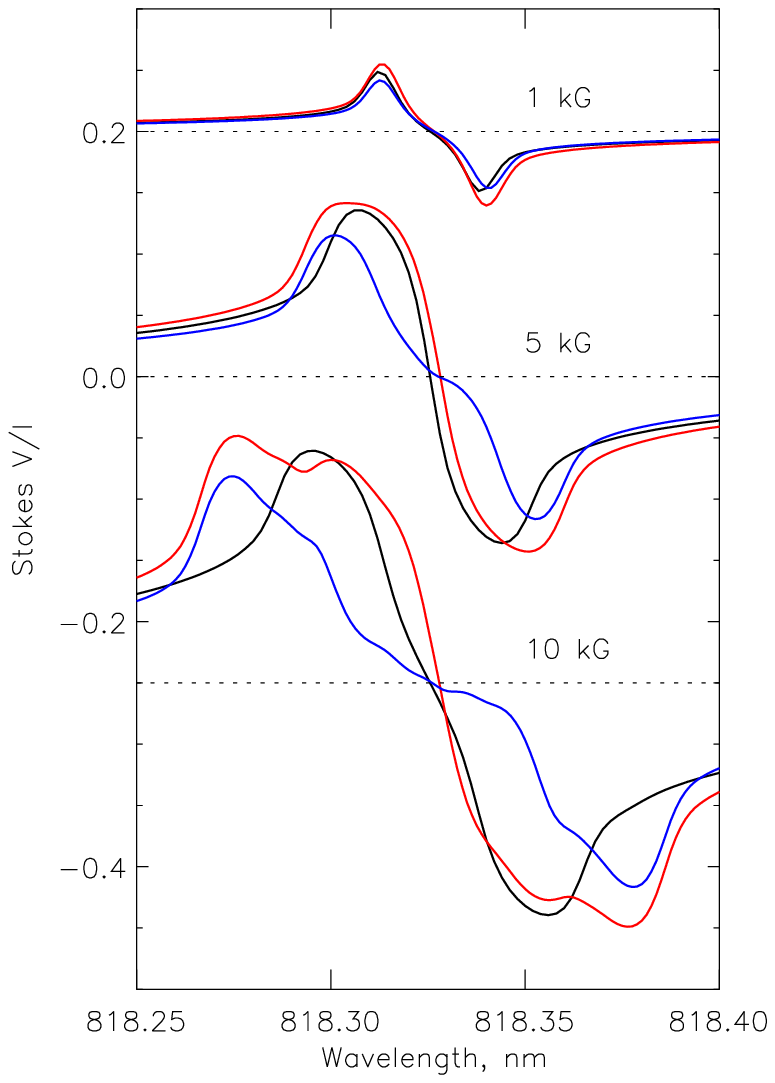}
\caption{
Stokes profiles of the Na I 818.3 nm line at magnetic field strengths 
of 1, 5, and 10\,kG. The profiles computed in the ZR assuming 
the field inclination angle $\gamma=0$ are shown with black lines.
The profiles computed in the PBR assuming $\gamma=0^\circ$ and 40$^\circ$ 
are shown with red and blue lines, respectively.
The profiles are shifted in vertically for clarity. 
The corresponding zero levels are shown with dotted lines.
The difference between the PBR and ZR profiles and the sensitivity to the
$\gamma$ angle due to the magneto-optical effect
increase with the field strength.
}
\label{fig:mod_sto}
\end{figure}

Example synthetic Stokes $V/I$ profiles of one Na I line are shown 
in Fig.~\ref{fig:mod_sto} for three magnetic strength values. To emphasize the
PBR effect, we also show profiles computed in the ZR, i.e. by
neglecting level mixing. As expected, the difference 
is especially noticeable at stronger fields. It is important
to notice that the line shape also varies and cannot be approximated
using the weak-field assumption. In fact, shapes of the Na I Stokes $V/I$
profiles and their relative amplitudes provide robust constraints 
on the field strength. If Stokes $V/I$ in the PBR is measured with a high
signal-to-noise ratio (and preferably with a high spectral resolution), 
it can uniquely identify the field strength, independently on the magnetic
field filling factor. We employ this sensitivity of the NIR Na I lines
in this paper.

\subsection{Magnetic Field on LSR J1835+3259}\label{sec:mod_lsrj}

To infer properties of the  magnetic field on LSR J1835,
we first fit the observed average Stokes $V/I$ profile 
shown in Fig.~\ref{fig:na_prof} with the synthetic PBR profiles by varying
the magnetic field strength, $B$, and its filling factor, $f$ (i.e., a fraction of the visible 
surface with the magnetic field) under two assumptions: 
(1) the field is homogeneous and longitudinal, i.e., along the line of sight, and 
(2) the field is homogeneous and inclined to the line of sight with the angle $\gamma$. 
These assumptions make no attempt to constrain the global field or satisfy div$B=0$.
Also, the field direction within the active regions may be neither longitudinal 
nor homogeneous, if the emission loops in Fig.~\ref{fig:em_cont} follow magnetic field lines.
Nevertheless, the first case provides the lowest limit for the magnetic field strength, 
and the second one can indicate more realistic estimates.
The modeled PBR Stokes $V/I$ for a given field 
strength and the angle $\gamma$ 
is simply scaled with the factor $f$: $V/I(B,\gamma,f) = f\times V/I(B,\gamma,f=1)$. 
In the cases (1) $\gamma=0^\circ$ or 180$^\circ$.
The magnetic region was assumed to be of the same temperature as the non-magnetic atmosphere. 
A change of the spot temperature would affect the filling factor but we do not 
expect large variations for cool dwarfs \citep[e.g.,][]{berdliv05}. 

Since the Na I Stokes profiles are in the PBR at $B>200$G, they have sensitivity
to disentangle the magnetic field strength and the filling factor for kG fields, 
even though this is limited by the SNR of the data. 
Also, in a strong-field regime, the magneto-optical effect introduces a polarization 
reversal in the middle of the Stokes $V$ for an inclined field.
This allows us to also infer the angle $\gamma$.
Examples of such profiles are shown in Fig.~\ref{fig:mod_sto}.
Hence, we obtain a range of feasible solutions for the three parameters: $B$, $\gamma$, and $f$. 

We have found that the field inclination angle $\gamma$ is well constrained by the Stokes $V$
profile shapes. For $B\ge1$\,kG (i.e., when the magneto-optical effect is strong), 
we obtain $\gamma=130^\circ\pm10^\circ$. Correspondingly, for the longitudinal field model, 
we find $\gamma=180^\circ$. Hence, the magnetic field is directed from the observer 
and most probably inclined with the angle of 130$^\circ$. This agrees
very well with the projected direction of the loops as illustrated in Fig.~\ref{fig:em_cont}.
The azimuth of these loops can characterize the azimuth angle $\chi$ of the $B$-field.

\begin{figure}
\centering
\includegraphics[width=8cm]{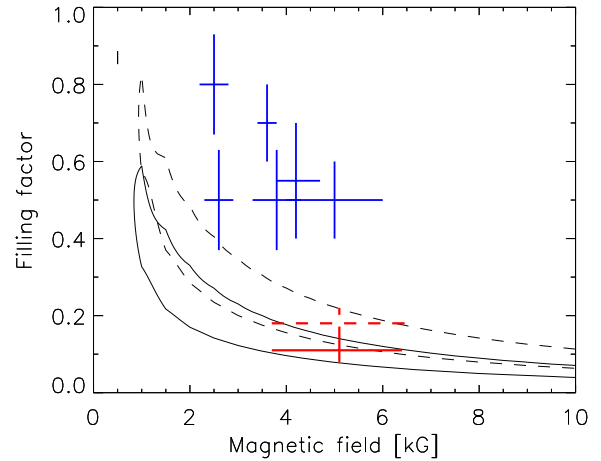}
\caption{
Best-fit magnetic field strengths and filling factors for the LSR J1835 average 
Stokes $V/I$ profile shown in Fig.~\ref{fig:na_prof}. 
The solid-line and dashed-line contours include feasible solutions for
a longitudinal ($\gamma=180^\circ$) and inclined ($\gamma=130^\circ$) fields,
respectively.
The red solid and dashed crosses mark the best solutions within the two contours,
which are compatible with the area of the active region recovered from emission lines.
Blue crosses indicate field parameters measured on M dwarfs 
\citep{jkv96,berdliv05} within their uncertainties. 
}
\label{fig:mod}
\end{figure}

In Fig.~\ref{fig:mod} we show envelopes of ($B$,$f$) solutions as contours for 
$\gamma=180^\circ$ (solid contour) and $\gamma=130^\circ$ (dashed contour).
Two immediate conclusions are that the field must be stronger than 1\,kG
and its filling factor must be $\ge4\%$.

We can further constrain the field strength in the active region near the
phase 0.0 on 2012 August 22 using its area from the Na I D emission contours
as shown in Fig.~\ref{fig:em_cont}. For instance, the 0.2 level of the maximum emission 
filling factor encompasses most of the emission flux. For the phases 0.0 to 0.2, 
during which the Stokes $V/I$ profile in Fig.~\ref{fig:na_prof} was measured and averaged, 
the area within this level is averaged to 11\%\ of the visible projected surface area. 
Now, according to the contours in Fig.~\ref{fig:mod}, the longitudinal field in this region 
is $5.1\pm 1.3$\,kG. The error on the filling factor from the same contour is
$\pm0.04$. An inclined field of the same strength requires $f=0.17\pm0.04$ or
up to 10\,kG for $f=0.11$. We note, however, that the error of $B$ increases for stronger
fields. Higher SNR and spectral resolution are needed for obtaining tighter constraints 
on the field parameters. 
The Stokes $V/I$ profile corresponding to the $B$=5.1\,kG and $f$=0.11 is shown with the 
red line in the lower panel of Fig.~\ref{fig:na_prof}.

We emphasize that this is the first detection of a surface magnetic field on a brown dwarf,
which is so far the coolest known dwarf with such strong magnetic fields. 
Earlier attempts to achieve this were inconclusive. For instance, an extensive study 
of magnetic properties of a sample of M7--M9 dwarfs by \cite{rb10} included only old
dwarfs with very modest activity levels. As as result, only a couple of objects were reported to have
magnetic fields of 3-4\,kG with an approximate error of $\pm$1\,kG. However, their analysis was 
based on fitting observed unpolarized flux spectra using a much warmer M3.5 spectrum of EV Lac
as the template. This approach leads to unknown biases in the value of the magnetic 
field strength and its error. In contrast, our approach to observe polarized spectra and
carry out detailed polarized radiative transfer taking into account numerous blends provides  
definite detection and realistic magnetic field parameters.

For EV Lac, we have assumed the field strength of 3.8\,kG \citep{jkv96} and obtained 
the best fit with the filling factor $f=28$\%\ and $\gamma=0^\circ$ for a longitudinal field
(Fig.~\ref{fig:evlac}) and $f=43$\%\ and $\gamma=50^\circ$ for an inclined field. 
The second value agrees well with the earlier estimate of the magnetic 
filling factor of $50\pm13$\%\ \citep{jkv96}. Parameters for a few very active 
red dwarfs (including EV Lac) are plotted in Fig.~\ref{fig:mod} for comparison 
\citep[from][]{jkv96,berdliv05}.

\section{Discussion}\label{sec:dis}

The observed circular polarization signature most probably represents a residual of multiple 
opposite-sign contributions from a complex field, which we may never be spatially resolve
(as on red dwarfs),
so we can interpret our measurements only in terms of residual homogeneous fields. 
However, modeling the PBR polarization in the Na I lines of LSR J1835 clearly indicates 
that the magnetic field in its active region should be stronger than 1 kG (Fig.~\ref{fig:mod}). 
An equipartition estimate 
of about 5\,kG can be inferred from the LSR J1835 model atmosphere assuming that the magnetic 
field strength should scale with the local gas pressure in the photosphere. 
In the solar atmosphere this value 
is about 1.5\,kG, and such fields are observed as small-scale magnetic elements in hot active 
regions (plages), while cool sunspots can harbor  a factor of two to three stronger fields, i.e, 
3--4.5\,kG.
On EV Lac, the equipartition magnetic field is 3.1\,kG at the base of its atmosphere, 
which is a bit lower than the observed 3.8$\pm$0.5\,kG. 
Our circular polarization measurements imply that a 5.1$\pm$1.3\,kG 
longitudinal field on LSR J1835 could occupy 11$\pm$4\%\ of the visible hemisphere 
and an inclined field would occupy 17$\pm$4\%. 
This is a factor of two to three smaller than inferred on early red dwarfs (e.g., EV Lac;
Fig.~\ref{fig:mod}). Whether this is related to a smaller magnetic flux or a larger field complexity
is to be clarified from high-resolution spectropolarimetry and MHD simulations.

Our measurements were carried out one month after a campaign of simultaneous radio and optical 
observations of LSR J1835 using Jansky Very Large Array (VLA) and LRIS spectrometer at Keck, 
respectively \citep{hal15}. Those observations confirmed the correlation of the optical 
emission with radio bursts. Also, it was suggested that this phenomenon is somewhat similar 
to auroras in planetary atmospheres bombarded by stellar wind particles. 
Our detection of the circular polarization signature due to a strong magnetic 
field near the maximum of the optical emission indicates that radio, optical, and magnetic signals 
originate from the same region which extends from the visible surface to higher layers above.

We can obtain indirect estimates of the magnetic field in the radio emission 
source regions, assuming that they radiate as an ECM.
This mechanism implies that radiation is emitted at the electron cyclotron frequency 
$\nu_{\rm c}\approx 2.8\times B$ MHz, 
where $B$ is the magnetic field strength in Gauss \citep[e.g.,][]{hal07}.
We obtain $B$ values along the loops L1 and L2 as shown in Fig.~\ref{fig:em_cont} (right panels):
1.5--1.6\,kG above H$\gamma$, 1.5--1.9\,kG within the H$\gamma$ regions, and 
2.5--2.8\,kG within the H$\beta$ regions. These loops are anchored at the magnetic
spot with 5.1\,kG in the photosphere.

\citet{hal15} estimated the bursting region to occupy only a few percent
of the visible dwarf surface. Our models indicate that the field strength 
within such a small region could be stronger than 10\,kG. 
Whether such a strong field can exist on the
brown dwarf surface remains to be clarified. An independent analysis of 
Stokes parameters within molecular bands, primarily in the CrH 860--870\,nm band,
which is also in the PBR, supports our detection of the 5\,kG field near 
the phase 0.0 \citep{kuzm16}. Also, the fact that the field is directed from the observer, 
with the inclination angle of about 130$^\circ$ agrees with the estimate from the CrH band 
of about 115$^\circ$ near phase 0.0. This further supports 
the recovered geometry of the magnetic emission loops, which are significantly 
inclined to the line of sight. Interestingly, the CrH polarization was marginally 
seen also at other rotational phases with smaller filling factors, 
which indicates that magnetic field may be small-scale and ubiquitous.

Different triggers can be considered for the strong and lasting bursts on LSR J1835 
and similar brown dwarfs. They do resemble planetary auroras, 
as was suggested by \citet{hal15}, when the planet magnetosphere is 
perturbed by the near-planetary environment, e.g., by stellar wind, as in the case of the Earth, 
or a moon's volcanic activity, as in the case of Jupiter and Io in the solar system.
An auroral nature of the bursts on LSR J1835 requires electron beams driven by 
large-scale current systems. Then, the emission will vary with variation of the current system.
This is possible if, for example, LSR J1835 would host a planet with a magnetosphere.
In the light of the recent discovery of the Proxima b around the M5.5 dwarf \citep{proxb} and 
the M8.5 dwarf TRAPPIST-1 system with seven planets on closely packed orbits 
\citep{trap1}, such "star--planet" interactions could power auroral activity on 
ultra-cool and brown dwarfs and their planets.

On the other hand, the relatively young age of LSR J1835 of 22\,Myr implies that it
should still possess some disk matter \citep{bar17} 
and perhaps continue accretion via magnetospheric accretion,
as proposed for T Tau stars \citep{hart94,shu94}. 
Observations of young clusters and associations 
suggest that dissipation of disks is less efficient for brown dwarfs 
than for higher-mass stars \citep[e.g.,][]{ria12,ria14,dow15}. 
In low-mass stars (0.1--1\,$M_\odot$), the transition from primordial disks 
to debris disks occur at about 10\,Myr, and debris disks peak at the age of
10-30\,Myr. In contrast, brown dwarf disks are still in the primordial stage
at 10\,Myr, and their debris disks seem largely disappear by the age of 40-50\,Myr
\citep{ria14}, with rare exceptions \citep{bou16}. 
Thus, it is possible that LSR J1835 still has a disk.
Interestingly, the observed IR spectral energy distribution (SED) of the dwarf
shows an excess which requires a significantly cooler
atmosphere with $T_{\rm eff}$=2200\,K and warm dust (BT-Settl models) 
than deduced from spectral lines \citep{kuzm16}.
Whether this is a consequence of the dust in the dwarf atmosphere or in a disk
or can be due to the presence of cool starspots
(or a combination of the three) has yet to be clarified. In particular, broadband polarimetry
can help distinguish between dust in a disk and in the atmosphere, while high-resolution
spectropolarimetry is needed to improve parameters of starspots.

The presence of a disk may facilitate large-scale
magnetic reconnections between the dwarf magnetosphere and the disk, 
similar to T Tau-type stars, as mentioned above \citep[see also][]{zhu09}. 
In this scenario, the strong stellar
magnetic field truncates the inner edge of the disk at the
distance where viscous ram pressure equalizes the magnetic
pressure. In pre-MS stars, this magnetospheric radius may reach
5--10 stellar radii. Inside this radius, the matter sporadically flows 
along magnetic field lines until it impacts
the stellar surface and causes bursts (flares).
Magnetic loops on brown dwarfs (as observed in the radio) may extend 
one to three dwarf radii from the surface \citep{lyn15}.
Thus, it is possible that radio-bursting ultra-cool dwarfs are in the T Tau 
phase of magnetospheric accretion. It would be necessary to verify whether they
possess kG-strong magnetic fields and are of young enough age, similar to LSR J1835.

In addition, larger filling factors for the equipartition field of 5\,kG 
(i.e., the most probable field strength for given plasma conditions) 
require wide-spread mixed-polarity magnetic fields similar to what is observed on the Sun 
as a small-scale, intergranular network field. 
This field constantly emerges from the turbulent interior, 
as is observed on the Sun and is predicted for fully convective ultra-cool dwarfs 
to be dominated by a highly entangled field 
at the equipartition level throughout the atmosphere. Then, this complex, rather strong 
field can also trigger explosive events. 
Such a complex field 
would frequently rearrange itself through reconnections leading to flares, 
aurora-like emission, and radio pulses, modulated by fast rotation of brown dwarfs. 
A combination of such surface magnetic field evolution with a possible dwarf--disk or 
dwarf--planet interactions would be an exciting opportunity to learn about early stages of the
planet formation in the presence of large-scale magnetosphere interactions.

\section{Summary and Conclusions}\label{sec:sum}

We have measured near-infrared polarized (Stokes $IQUV$) and optical emission spectra 
of the active M8.5 young brown dwarf LSR J1835+3259 using the LRISp spectropolarimeter 
at the Keck telescope during two consecutive nights on 2012 August 22-23.
Measurements were made at several aspect angles during two rotational periods separated
by seven periods. 
The 5.1\,kG magnetic field was detected as a Zeeman signature in the near-infrared Na I 819\,nm 
lines during the first night when an active region radiating non-thermal optical emission faced 
the Earth (near phase 0.0). The magnetic field is found to cover at least 11\%\ of the dwarf 
visible hemisphere.

By employing the sensitivity of the NIR Na I and TiO lines in the 819\,nm region 
to the temperature and gravity, we have determined atmospheric parameters 
$T_{\rm eff}$=2800$\pm$30\,K and $\log g$=4.50$\pm$0.05 with high accuracy. 
A comparison with evolutionary models leads us to the conclusion that 
LSR J1835+3259 is a young brown dwarf with the mass $M=55\pm 4 M_{\rm J}$ 
and age $t=22 \pm 4 {\rm Myr}$. Thus, its prominent magnetic activity is probably related
to its young age.

We have inferred the emitting region topology using optical emission line profile inversions.
We have found that model line profiles corresponding to a decaying stage of an energetic flare 
on red dwarfs are well suited for such inversions. The emission maps obtained from inversions
indicate the presence of hot plasma loops of at least 7000\,K with a vertical stratification 
of the emission sources. In particular, it was found that higher-excitation emission peaks
at earlier phases than the lower excitation emission.
As the dwarf rotates with the 2.84\,hr period, these loops rotate in and out of view and cause 
the emission modulation observed as periodic bursts. The emission near the phase 0.0 was the strongest 
on 2012 August 22, and decayed by 2012 August 23, i.e., on the time scale of several rotation periods, 
making such periodic bursts transient. At the same time, a second emission region was emerging 
at almost opposite longitude (phase 0.6) toward the end of our observing campaign.

The region near the phase 0.0 was also active (in optical and radio emission) 
a month before our measurements \citep{hal15}. Hence, the life time of this active region 
is at least a month, and at least two transient flare-like events occurred in this region. 
It was also puzzling that radio pulses were arriving to Earth at somewhat earlier phases 
than H$\alpha$ emission. Our emission maps interpreted as projections of extended corotating 
loops indicate that the high-frequency radio emission sources appear co-spatial with high-excitation
optical emission (perhaps due to projection).

The 5\,kG magnetic field was detected at the base of the emission loops near the phase 0.0. 
This is the first time that we can quantitatively associate radio and optical bursts with a strong, 
5\,kG surface magnetic field required by the ECM instability mechanism.
When assuming the ECM mechanism, the magnetic field strength along the loops seems reduce 
from 5\,kG at the base to 1.5\,kG at the highest observed location, which can be as high as 
three dwarf radii from the surface \citep{lyn15}. Whether
this magnetic region is associated with the global magnetic field of the dwarfs is an interesting 
question. A longer series of polarimetric observations is needed to verify this. Also, high-resolution
full-Stokes spectropolarimetry can help to map magnetic field and emission regions on this dwarf 
with a better spatial resolution. 

We conclude that the activity on LSR J1835+3259 and possibly other ultra-cool and brown dwarfs 
with non-thermal radio and optical emission bursts is associated with a strong magnetosphere driven by  
a few kG surface magnetic fields, similar to what is observed on active red dwarfs.
An interaction of a large-scale magnetic field (long-lived active regions or magnetic poles) 
with small-scale, entangled, widespread and rapidly evolving magnetic fields and
possibly with a magnetized disk or a planet (to be confirmed) may lead to
frequent reconnection events and trigger optical and radio bursts.
Since this is the coolest known dwarf with such a strong magnetic field,
our result also provides a unique constraint for simulations of magnetic fields 
in fully convective ultra-cool dwarfs \citep[e.g.,][]{dob06,brow08,yad15}. 
Considering that hot Jupiter-like exoplanets are of similar temperatures 
as brown dwarfs, our result brings us closer to studying the magnetism of hot Jupiters.
\\

This work was supported 
by the ERC Advanced Grant HotMol (www.hotmol.eu) ERC-2011-AdG-291659.
Based on observations made with the Keck Telescope, Mauna Kea, Hawaii.
We thank the Keck staff, support astronomers and, in particular,
Dr. Bob Goodrich and Dr. Hien Tran for their support.
S.V.B. acknowledges the support from
the NASA Astrobiology Institute and the Institute for Astronomy,
University of Hawaii, for the hospitality and allocation
of observing time at the Keck telescope.
The authors wish to recognize and acknowledge the very significant 
cultural role and reverence that the summit of Mauna Kea has always had 
within the indigenous Hawaiian community. we are most fortunate to have 
the opportunity to conduct observations from this mountain.
We thank an anonymous referee for the constructive and helpful report.

\end{document}